\documentstyle[preprint,epsf,aps]{revtex}
\begin{document}
\title{ Producing the entangled photon pairs without type II down conversion}
\author{Wang Xiang-bin\thanks{email: wang$@$qci.jst.go.jp} 
\\
        Imai Quantum Computation and Information project, ERATO, Japan Sci. and Tech. Corp.\\
Daini Hongo White Bldg. 201, 5-28-3, Hongo, Bunkyo, Tokyo 113-0033, Japan}

\maketitle 
\begin{abstract}
We propose a simple scheme to produce the polarization entangled photon pairs without the type II phase match.
The same scheme can also be used to produce the macroscopic entangled photon states in both photon number space and
the polarization space( A. Lamas-Linares, J.C. Howell and D. Bouwmeester, Nature, 412887(2001)). Advantages and 
applications of our scheme in quantum key distribution are discussed. 
\end{abstract}
As it is general believed that, quantum entanglement plays a fundamentally important role in the quantum non-locality\cite{bell},
quantum information and quantum computation(see e.g., \cite{pan1,bou,ekert,tittel,ekert2}). Entanglement photon pairs have been regarded as a type of
 important entanglement
resource in the past study towards the quantum entanglement. It seems that the type II parametric down-conversion is the 
best way to produce the entangled photon pairs. As it has been shown that, the type I parametric down conversion
will only produce correlated photon pairs with the same polarization, however, the type II conversion-down process
will emit the true entangled photon pairs with certain small probability.
In this letter we propose a different simple scheme to produce the entangled photon pairs from two independent 
sources. Take as an example, we want to produce the following two photon Bell state
\begin{eqnarray}
|\Psi\rangle = \frac{1}{\sqrt 2}(|H\rangle_a|V\rangle_b-|V\rangle_a |H\rangle_b)\label{en}
\end{eqnarray}
where $H,V$ represent for the horizontal and vertical polarization respectively, the subscript $a,b$ represent for
the different subspaces(i.e., the different modes), for simplicity we shall obmit the subscripts and always imply that the first register is in subspace $a$
and the second is for the subspace $b$. The above defined entangled state is  the linear superposition of the state
$|H\rangle|V\rangle$ and the state $|V\rangle |H\rangle$.
Suppose we have two independent sources, 1 and 2.
Source 1 emits the states
\begin{eqnarray}
|\psi_1\rangle=\sqrt{1-P^2}|00\rangle+ P|H\rangle|V\rangle
\end{eqnarray}
and source 2 emits the state
\begin{eqnarray}
\psi_2\rangle=\sqrt{1-P^2}|00\rangle- P|V\rangle|H\rangle,
\end{eqnarray}
and $|P|$ is very small.
Actually, these sources can be very well approximated by two two mode weakly squeezed states sources. 
Suppose  $P$ is so small that $P^2$, the probability that both sources emit two photons are 
very small and such events will be ignored in our paper.
Under such circumstance, we only need to erase the source information in detecting the photons, i.e. we should
 arrange something
so that finally we only known that there are two photons there but we do not know where they come from.
The source information can be erased by the polarized beam splitters. A polarized beam splitter(PBS)  
reflects the vertically polarized photon and transmits the horizontally polarized photon(see fig. 1).
The scheme to make the indistinguishability is shown in Fig. 2.
Let the beams of mode $a$ of each source be incident to PBS 1, with one beam in one side of the PBS 1. 
The two beams are with the same angle with PBS 1.
Let the beams of mode $b$ of each source be incident to PBS 2, with one beam in one side of the PBS 1. 
The two beams are with the same angle with PBS 2.
In such an arrangement, if we took a measurement on the photon number of the total output and we obtain 2 photons as the measurement
 result ,
we get the entangled state as defined in eq(\ref{en}). Actually, this total photon number detection is not necessary
in many applications, such as the Bell inequality testing\cite{kwiat} and the quantum teleportation\cite{bou}.
Because we can exclude the no photon events  and only count in the cases that both detectors are fired. 

If we replace the weak squeezed states sources by the strong ones, we can obtain the nontrivial entangled state 
on both
photon numbers and the polarizations. Suppose now the state from each source is
\begin{eqnarray}
|\psi_1(r)\rangle = \sum_{k=0}^{\infty}\tanh^k r |k\rangle_h|k\rangle_v
\end{eqnarray}
and
\begin{eqnarray}
|\psi_2(r)\rangle = \sum_{k=0}^{\infty} 
\tanh^k (-r)|k\rangle_v|v\rangle_h.
\end{eqnarray}
where the subscript $h,v$ represent the horizontally or vertically polarizion respectively, 
the real number $r$ is the squeezing parameter.
With such sources, from the same scheme in Fig.2 we may obtain the following state
 \begin{eqnarray}
|\Phi\rangle = \sum_{n=0}^\infty \tanh^{n}r \sum_{m=0}^{n} 
(-1)^m(|n-m\rangle_h|m\rangle_v)_a (|n-m\rangle_v|m\rangle_h)_b.
\label{la}\end{eqnarray}
A similar state has been created already by the type II down-conversion experiment\cite{lamas}. 
If we took a measurement on the total photon number in either mode, we would obtain a maximally entangled
state as
\begin{eqnarray}
|E_n\rangle =\sum_{m=0}^{n} (-1)^m (|n-m\rangle_h|m\rangle_v)_a (|n-m\rangle_v|m\rangle_h)_b.
\end{eqnarray}
As it has been shown in\cite{durkin}, since the state $|E_n$ is invariant under a rotation to the polarization
by $\pi/4$ angle,  such a state can be used to carry out  the quantum 
key distribution(QKD) task. If we can produce the state with large $n$, the QKD using such a state has an advantage or higher
bit rate than that of the one using the single polarized photons.
Actually, in doing the quantum key distribution, one can directly use the state defined
in eq(\ref{la}) and then take a post selection.
Suppose after produced the state $|\Phi\rangle\rangle$ Alice sends the mode $b$ to Bob and keeps mode $a$.
They each will then independently and randomly choose a basis from $H/V$ or $\frac{\pi}{4}/-\frac{\pi}{4}$ and take 
the measurement. If they have chosen the same basis, their measurement result should be correlated, i.e., whenever
Alice obtains  $n-m$ photons on certain polarization and $m$ photons on the orthogonal polarization, 
Bob should have a result of  $m, n-m$ photons on the corresponding polarization. 
The effect of losses and the security are also studied in\cite{durkin}.  
As it has been pointed out, the main advantage of this quantum key distribution scheme is in its high efficiency. 
Compared with the traditional quantum key distribution schemes, here the bit
rate rises significantly with the number of photons $n$ because the number of distinguishable results is increased.
\\
Unfortunately, if the state $|\Phi\rangle$ is prepared by the type II parametric down-conversion, the above mentioned
 advantage is not so significant, because the probability to collapse to a large $n$ state is rather small, as
it was shown in ref\cite{lamas}. In the type II parametric down-conversion scheme,  the different
mode beams diverge geometrically, and the horizontal and vertical polarized photons experience different crystal parameters
. So the useful crystal length is limited\cite{ata} thereby any efficient emission is prohibited.
Due to this fact, the quantum key distribution scheme proposed in\cite{durkin} has little advantage if a source of type II
parametric conversion-down is used.
However, our scheme does not require a type II parametric conversion-down source and there is in principle 
no limitation to the value of squeezing parameter $r$ in our source. This opens the chance to produce the state 
$|\Phi\rangle$ of which large $n$ states contribute signinificantly.
{\it This is to say that by our scheme, 
the optical entangled states with large photon number can be created and the efficient quantum key distribution
scheme\cite{lamas} can be done.}
A schematic diagram of the efficient QKD scheme is shown in Fig. 3.
 
In summary, we have proposed a simple scheme to produce the entangled photon states by using the polarized beam splitters. 
The scheme can have important
advantage in doing the quantum key distribution.  
Previously\cite{pan2}, it was proposed to make the entanglement entanglement concentration by using PBS. It seems that polarized beam 
splitters will have an important role in distilling or creating the quantum entanglement. 
I thank Prof Imai for support. I thank Dr Tomita A, Dr Fan H, Dr Matsumoto K and Dr Yura A  for useful
discussions.

\begin{figure}
\begin{center}
\epsffile{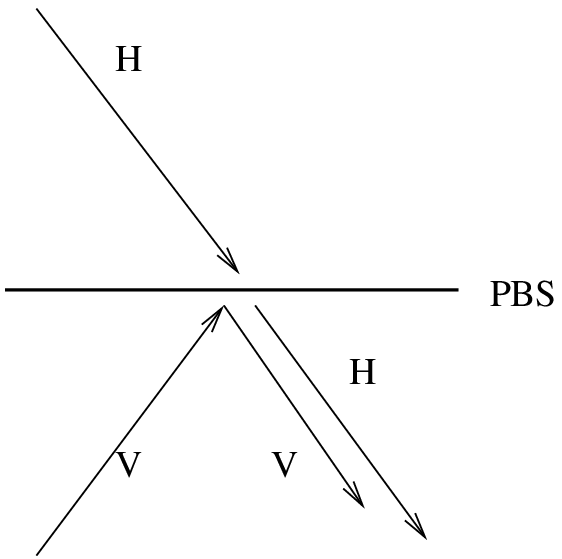}
\end{center}
\caption{ A schematic diagram for the property of a polarized beam spliter(PBS). 
It transmits a horizontally polarized
photon $H$ and reflects a vertically polarized photon $V$.} 
\end{figure}
\newpage
\begin{figure}
\begin{center}
\epsffile{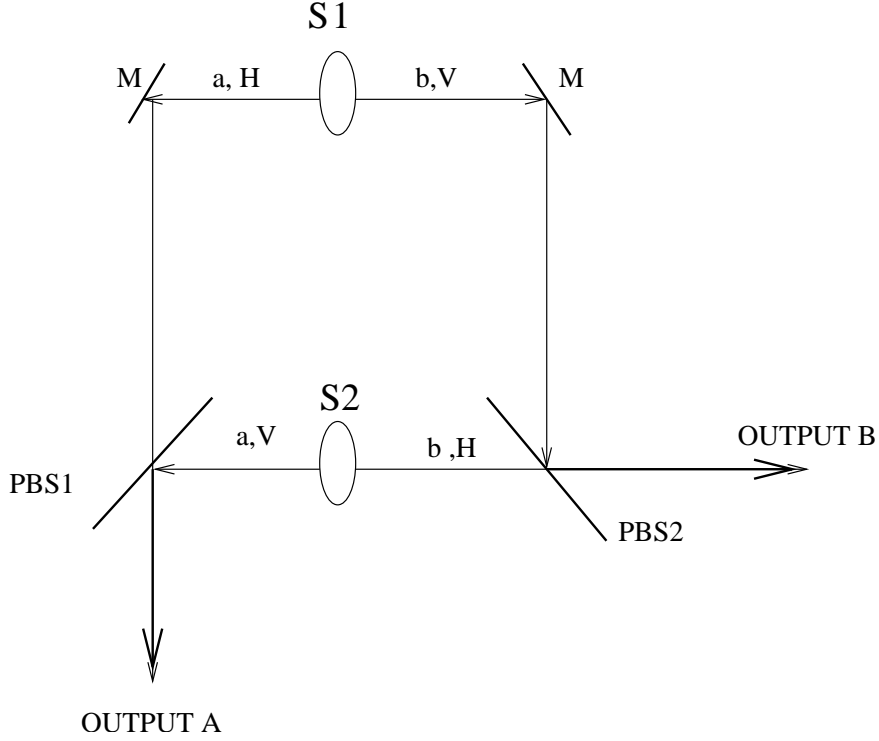}
\end{center}
\caption{ Polarized beam splitters can help to produce the entangled states. M  represents for mirrow. 
There are two independent sources
S1 and S2, each emits two mode squeezed states $|\psi_{1}\rangle =\sum_{k=0}^\infty\tanh ^k r |k\rangle_H|k\rangle_V $ 
and $|\psi_{2}\rangle =\sum_{k=0}^\infty\tanh ^k r |k\rangle_V|k\rangle_H $ respectively.
In the case $r$ is small, to a good approximation, an entangled pair in mode A and B is obtained in the case that 
there are
two photons in the output. This suffices to for the observation of Bell inequality violation. In the case that $r$ is large,
the scheme offers  photon number and polarization entangled states with large photon number. 
Type II parametric conversion-down process can only
offer such  a state with small photon number. }
\end{figure}

\begin{figure}
\begin{center}
\epsffile{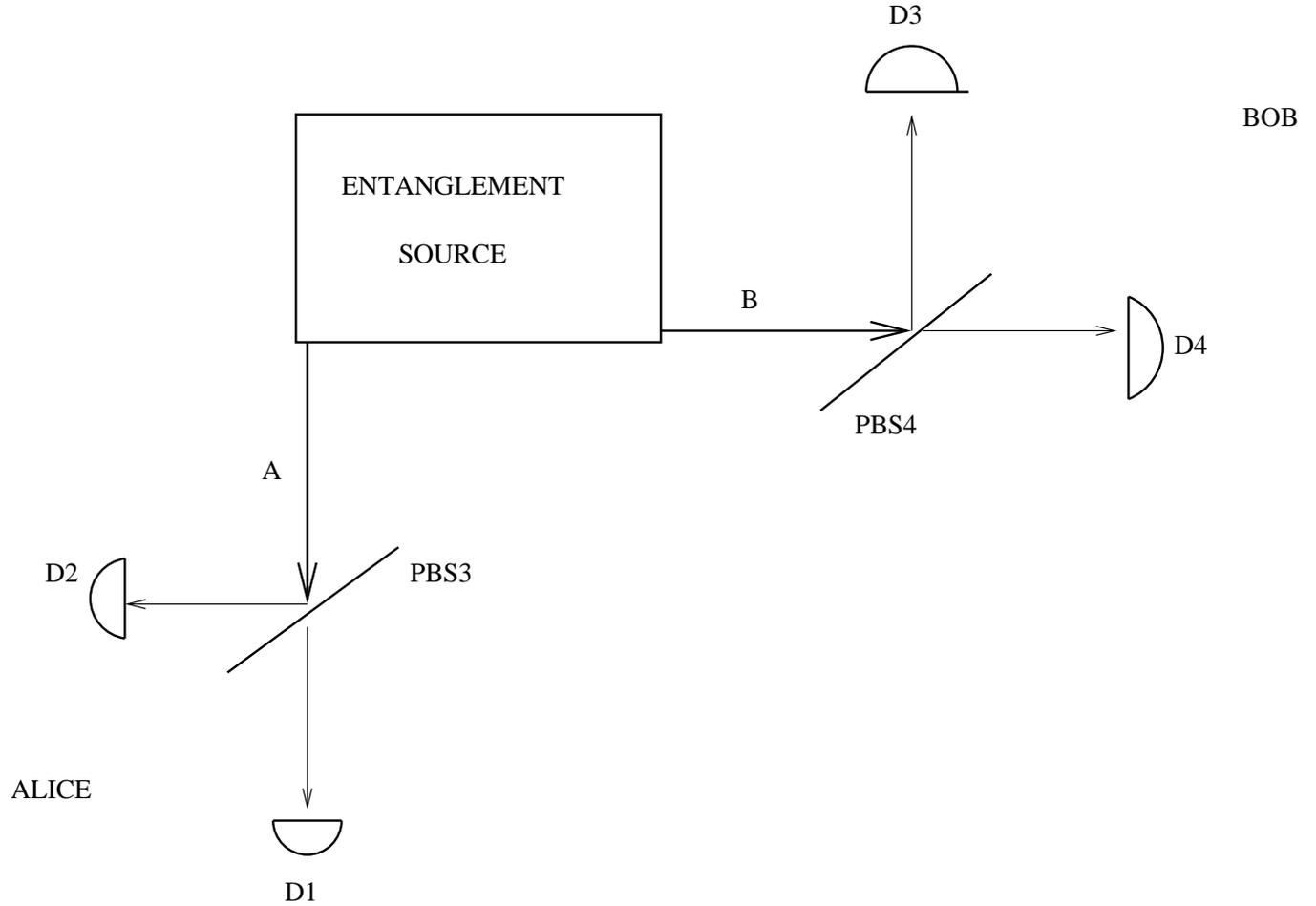}
\end{center}
\caption{ An efficient scheme for quantum key distribution.
The entanglement source is the source by the scheme from Fig.2. D1, D2, D3 and D4 are photon detectors. 
They can measure the photon numbers in each of the two orthogonal polarization 
exactly. Note that each time before detecting beam A by the detectors D1 and D2, 
Alice  randomly chooses an
$H/V$ PBS( a PBS that transmits $H$ polarization and reflects $V$ polarization) or a $\frac{\pi}{4}/ -\frac{\pi}{4}$(
a PBS that transmits $\pi/4$ polarization and reflects $-\pi/4$ polarization ) as her PBS3. Similarly, 
Bob also randomly and independently
chooses $H/V$ or   $\frac{\pi}{4}/ -\frac{\pi}{4}$ PBS as his PBS4 before 
he measures the photons on beam B by detectors $3$ and $4$.} 
\end{figure}
\end{document}